# The peculiar outburst activity of the symbiotic binary AG Draconis

R. GÁLIS[1], J. MERC[1,2], L. LEEDJÄRV[3], M. VRAŠŤÁK[4] & S. KARPOV[5,6,7]

(1) Institute of Physics, Faculty of Science, P. J. Šafárik University in Košice,
Park Angelinum 9, 040 01 Košice, Slovakia, rudolf.galis@upjs.sk

(2) Astronomical Institute, Faculty of Mathematics and Physics, Charles University,
V Holešovičkách 2, 180 00 Prague, Czech Republic

(3) Tartu Observatory, Faculty of Science and Technology, University of Tartu,
Observatooriumi 1, Tõravere, 61602 Tartumaa, Estonia

(4) Variable Star and Exoplanet Section of Czech Astronomical Society,
LSO (private observatory), 03401 Liptovská Štiavnica, Slovakia

(5) CEICO, Institute of Physics, Czech Academy of Sciences,
Na Slovance 1999/2, 182 21 Prague, Czech Republic

(6) Special Astrophysical Observatory, Russian Academy of Sciences,
Nizhniy Arkhyz 369167, Russia

(7) Institute of Physics, Kazan Federal University,
16a Kremlyovskaya St., Kazan 420008, Russia

**Abstract:** AG Draconis is a strongly interacting binary system which manifests characteristic symbiotic activity of alternating quiescent and active stages. The latter ones consist of the series of individual outbursts repeating at about a one-year interval. After seven years of flat quiescence following the 2006–2008 major outbursts, in the late spring of 2015, the symbiotic system AG Dra started to become brighter again toward what appeared to be a new minor outburst. The current outburst activity of AG Dra was confirmed by the following three outbursts in April 2016, May 2017 and April 2018. The photometric and spectroscopic observations suggest that all these outbursts are of the *hot* type. Such behaviour is considerably peculiar in almost 130-year history of observing of this object, because the major outbursts at the beginning of active stages are typically *cool* ones. In the present work, the current peculiar activity of the symbiotic binary AG Dra is described in detail.

## Introduction

AG Dra is one of the best studied symbiotic systems. Its cool component is a metal-poor cool giant of spectral type K3 and higher luminosity than that of standard class III (Smith et al., 1996). The hot component of AG Dra is probably a white dwarf (WD) sustaining a high temperature of $(1-1.5) \times 10^5$ K and luminosity of $(1-5) \times 10^3$ L$_\odot$ due to the thermonuclear burning of accreted matter on its surface (Mikołajewska et al., 1995; Sion et al., 2012). The giant is under-filling its Roche lobe and the accretion most likely takes place from the stellar wind of the cool giant. Both components are in a circumbinary nebula, partially ionised by the WD.

The period analysis of long-term photometric and spectroscopic observations confirmed the presence of two periods in AG Dra (Hric et al., 2014). The longer one ($\approx$551 d) is related to the orbital motion and the shorter one ($\approx$355 d) could be due to pulsation of the cool component in this symbiotic system (Gális et al., 1999; Friedjung et al., 2003). The orbital period is mainly manifested during the quiescent stages of AG Dra at shorter wavelengths (*U* band), while the pulsation period is present during quiescent as well as the active stages at longer wavelengths (*B* and *V* bands).

The period analysis of active stages confirmed the presence of a period of around 360 d, which is the median of the time interval between outbursts. It is worth noting that these time intervals vary from 300–400 d without an apparent long-term trend. Most of the longer periods (e.g. 1330, 1580, 2350, 5500 d) are more likely related to the complex morphology of the light curve (LC) during the active stages than to the real variability present in this symbiotic system (Hric et al., 2014).

The LC of AG Dra, available since 1890 (Robinson, 1969), manifests characteristic symbiotic activity with alternating quiescent and active stages (Fig. 1). The latter ones occur in intervals of 9–15 yr and consist of several outbursts repeating at about one-year interval with a brightening of about 1–1.4 mag in the *V*/visual band and up to 2.3 and 3.6 mag in the *B* and *U* bands, respectively. During the period 1890–2018, AG Dra underwent six (or seven?) stages of activity: A (1932–1939), B (1949–1955), C (1963–1966), D (1980–1986), E+F (1993–2008) and G (2015–). In total, we recognized 36 outbursts in this period.

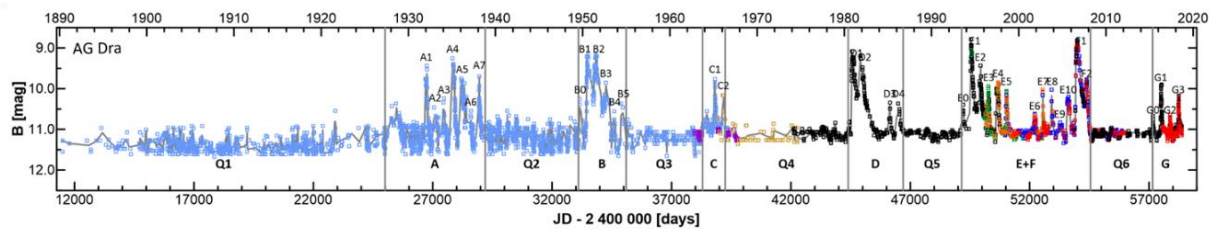

**Figure 1:** The historical LC of AG Dra, which manifests 129 yr of the photometric history of this symbiotic system, was constructed based on photographic (1890–1966; Robinson, 1969) and *B* band observations (1966–2018; Hric et al. 2014; this work). The LC is divided into active (A–G) and quiescence (Q1–Q6) stages by vertical lines. Particular outbursts are assigned as A1–A7, B1–B5, C1, C2, D1–D5, E0–E10, F1, F2 and G0–G3. The thin curves show spline fits to the data points.

UV and X-ray observations showed that there are two types of outbursts: *cool* and *hot* ones (González-Riestra et al., 1999). Major outbursts at the beginning of active stages (e.g. 1981–1983, 1994–1996 and 2006–2008) are usually *cool*, during which the expanding pseudo-atmosphere of the WD cools down and the He II Zanstra temperature drops. In smaller scale *hot* outbursts, the He II Zanstra temperature increases or remains unchanged. In the previous work, we demonstrated that the outbursts of AG Dra can also be clearly distinguished according to the behaviour of the prominent emission lines in optical spectra (Leedjärv et al., 2016).

The nature of these periodical outbursts has been a matter of long-term debate. One of the promising explanations of at least some individual outbursts of AG Dra might be the combination nova model proposed for Z And by Sokoloski et al. (2006). In this model, thermonuclear reactions are ignited when accretion rate onto the WD exceeds some critical value, and luminosity of the hot component increases significantly. One of the subsequent tasks would be to study whether the recent outbursts of AG Dra will fit into such a picture.

**Observations**

In this study, we analysed the photometric and spectroscopic observations that cover the ongoing active stage of AG Dra. The new photometric measurements were obtained during 167 nights at the Liptovská Štiavnica Observatory using the Newtonian telescope 355/1600 equipped with CCD G2-1600 and the set of Johnson–Cousins *U*, *B*, *V*, $R_C$ and $I_C$ filters. The AG Dra system was also observed for 358 nights using the 9-channel wide-field optical monitoring system with sub-second temporal resolution, Mini-MegaTORTORA, in operation at the Special Astrophysical Observatory of Russian Academy of Sciences in Caucasus. We also utilised the observations from *AAVSO International Database* (Kafka, 2018) acquired for 721 nights. To compare the behaviour of the latest outbursts of AG Dra with previous ones, we also used all photometric observations of this symbiotic system that had already been analysed and discussed in our previous study (Hric et al., 2014).

The spectroscopic observations of AG Dra were acquired by *ARAS Group*[2] observers mostly in the framework of two observing campaigns which we initiated and coordinated in 2017 and 2018. Although the spectra were obtained with small telescopes (25–35 cm, R ≈ 1800–11000), they provided us valuable information about the recent activity of AG Dra. In total, we used 278 spectra covering the time interval from JD 2 456 765 (April 17, 2014) to JD 2 458 447 (November 24, 2018). Moreover, we analysed the new intermediate-dispersion spectra of AG Dra obtained at the Tartu Observatory in Estonia (4 spectra, 1.5-m telescope, R ≈ 6000 and 7000) and at the Observatory of the Astronomical Institute of ASCR in Ondřejov (16 spectra, 2.0-m telescope, R ≈ 13000).

Our analysis was focused on the prominent emission lines in the wavelength regions under study: the hydrogen Balmer lines H$\alpha$ ($\lambda$ 6563) and H$\beta$ ($\lambda$ 4861), the neutral helium He I ($\lambda$ 6678) line, the ionised helium He II ($\lambda$ 4686) line, and the Raman-scattered O VI line at $\lambda$ 6825. Equivalent widths (EWs), fluxes in lines, peak intensities relative to the continuum and the positions of these lines were measured.

**Recent outburst activity of AG Dra**

After seven years of quiescence following the 2006–2008 major outbursts, the symbiotic system AG Dra started to become brighter again toward what appeared to be a new minor outburst in the late spring of 2015 (Fig. 2). The outburst activity of AG Dra was definitely confirmed by the following three outbursts in April 2016, May 2017 and April 2018 (Gális et al., 2018). In the next sections, the photometric and spectroscopic behaviour of the symbiotic system AG Dra during ongoing outburst activity stage are described in detail.

---

[2] http://www.astrosurf.com/aras/Aras_DataBase/Symbiotics.htm

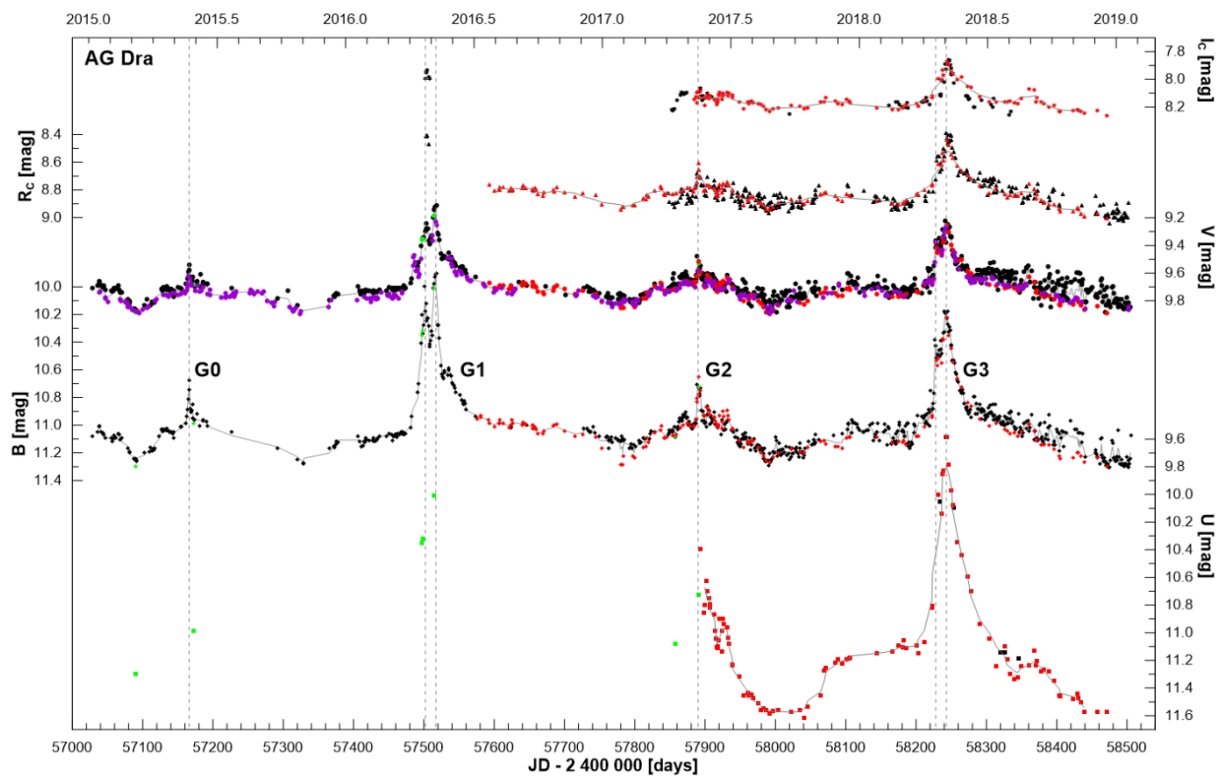

**Figure 2:** The LCs of the symbiotic system AG Dra during the recent active stage (2015–2018) constructed on the basis of the $U$, $B$, $V$, $R_C$ and $I_C$ band observations. The new photometric measurements are depicted by red and purple symbols for observations obtained at the Liptovská Štiavnica Observatory and the Special Astrophysical Observatory in Russian Caucasus, respectively. The green and black symbols represent observations acquired from Munari et al. (2015, 2016, 2017) and AAVSO (Kafka, 2018), respectively. The thin curves show spline fits to the data points. The dashed vertical lines indicate the times of individual brightness maxima of AG Dra during the ongoing active stage.

*Photometric behaviour*

The first, less prominent outburst (G0) was observed in May 2015. The maximum brightness was achieved around JD 2 457 166 (10.7 and 9.6 mag in the $B$ and $V$ bands, respectively). It turned out that it was minor outburst of AG Dra, a precursor of its activity as it was observed in some of the previous active stages. The major outbursts were preceded by the pre-outbursts with brightness around 10.4 and 9.4 mag in $B$ and $V$ band, respectively, in the case of the active stages B, E and probably C.

During the second, more prominent outburst (G1), the brightness around JD 2 457 517 (May 8, 2016) reached the maximum of 9.9 and 9.1 mag in the $B$ and $V$ bands, respectively. Actually, the outburst had double-peak structure with the first minor brightening occurring 15 days (JD 2 457 502) before the main one.

As in the case of previous outburst (G0), its amplitude ranks this brightening to the minor outbursts of AG Dra. Such photometric behaviour of the active stage is very unusual in the historical LC of AG Dra. More often, the pre-outbursts of AG Dra are followed by major outbursts, during which the brightness can reach around 8.8 and 8.4 mag in the $B$ and $V$ bands, respectively. A major outburst was not observed only during activity stage in 1963–1966, which was the shortest one in the almost 130-year photometric history of this interacting symbiotic system.

In May 2017, the third brightening (G2) during the recent activity of AG Dra was detected. It was a very sharp and short-lasting outburst of the *hot* type. Maximal brightness of the symbiotic system reached at JD 2 457 890 (10.7, 9.5 and 8.6 mag in the $B$, $V$ and $R_C$ bands, respectively) was similar to the case of G0. In the filter $I_C$, the outburst G2 was not detected at all. Very low amplitude of this brightening would be related not only to specific physical conditions during this outburst but also to orbital motion since the symbiotic system AG Dra was in the photometric minimum according to the orbital emphemeris given by Gális et al. (1999).

According to our statistical analysis of photometric observations, we determined that time intervals between outbursts of the symbiotic system AG Dra vary from 300 to 400 d (without an apparent long-term trend), with a median around 360 d. Therefore, we expected the next outburst in the interval from April 21, 2018 (JD 2 458 230) to May 31, 2018 (JD 2 458 270).

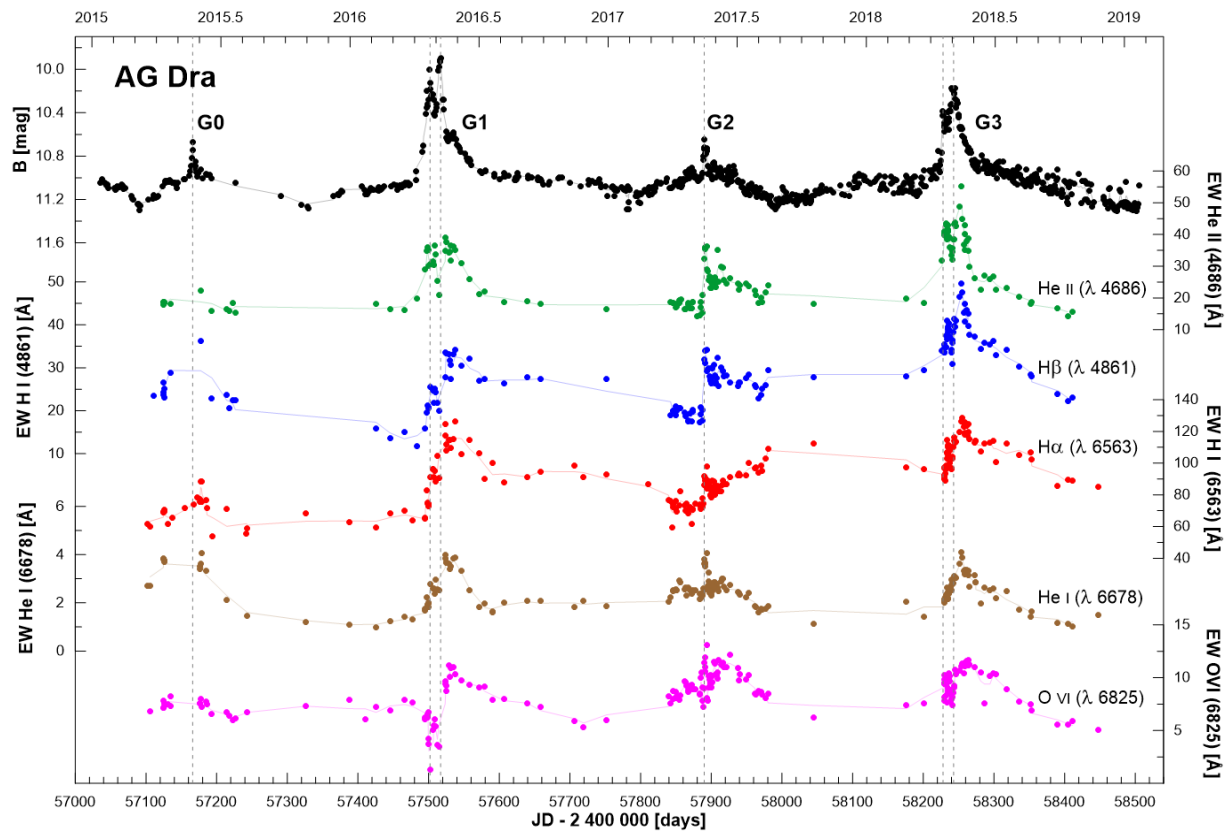

**Figure 3:** The light curve of AG Dra during the recent active stage in the *B* band together with the curves of studied emission line EWs. The thin curves show spline fits to the data points. The dashed vertical lines indicate the times of individual brightness maxima of AG Dra during the ongoing active stage.

Actually, the AG Dra system had already begun a rise in brightness, reaching 10.82, 10.95, 9.67, 8.77 and 8.13 mag in the *U*, *B*, *V*, $R_C$ and $I_C$ bands, respectively, on April 14, 2018. The outburst was confirmed on April 20, 2018 when the brightness of AG Dra increased to 10.53 and 9.42 mag in the *B* and *V* filters, respectively. We alerted the astronomical community (Gális et al., 2018) and initialised the observational campaigns to study photometric and spectroscopic behaviour of the recent outburst of AG Dra.

The maximum of the fourth outburst of the ongoing activity stage was reached 353 days after previous one, on May 4, 2018 (JD 2 458 243), with brightness of 9.6, 10.2, 9.2, 8.4 and 7.9 mag in the *U*, *B*, *V*, $R_C$ and $I_C$ bands, respectively. As in the case of the G1 outburst, the latest one had double-peak structure with the first minor brightening occurring around 25 days (JD 2 458 228) before the main one.

At the end of July 2018, the brightness of AG Dra almost returned to values typical for quiescence (11.4, 11.1 and 9.8 mag in *U*, *B* and *V* bands, respectively), so the fourth outburst has finished (Merc et al., 2018). The photometric behaviour suggests that all four recent outbursts of AG Dra belong to the minor, *hot* type. Such classification is also supported by the increase of the EWs of studied emission lines detected during all these events.

*Spectroscopic behaviour*

We analysed variability of selected emission lines in the optical spectrum of AG Dra during almost 14 years (1997–2011) using own intermediate-dispersion spectroscopic observations (Leedjärv et al., 2016). It is worth noting that these emission lines originate in the circumbinary nebula, which is generated by the stellar wind of the cool giant. Moreover, the nebula is partially ionized by short-wave radiation of the WD, resulting in its complex structure and variability. We studied the variability of EWs of the selected emission lines: H$\alpha$ ($\lambda$ 6563), H$\beta$ ($\lambda$ 4861), He I ($\lambda$ 6678), He II ($\lambda$ 4686) and the Raman-scattered O VI line at $\lambda$ 6825.

One of the most interesting features of this variability is the significant increase of the EWs of all the five emission lines considered, but in particular that of H$\alpha$ and Raman-scattered O VI ($\lambda$ 6825), during some minor outbursts of this symbiotic system (e.g. E10). On the other hand, the major, *cool* outbursts of AG Dra (e.g. in July 2006) are not specifically distinct in the EWs of hydrogen and helium lines, but the weakening of the Raman-scattered O VI ($\lambda$ 6825) line is very well seen.

A simple interpretation of this behaviour could be that during the *cool* outburst, the temperature of the hot component of the symbiotic system decreased considerably, so that the high excitation Raman-scattered O VI (λ 6825) line faded significantly and almost disappeared, however leaving the lower excitation emission lines of hydrogen and helium mainly unaffected.

Direct comparison of the spectra of AG Dra obtained during the quiescence stage Q6 (JD 2 456 906) and the pre-outburst G0 (JD 2 457 176) reveals significant increase of the EWs of all studied emission (Fig. 3). Such spectroscopic behaviour is typical for the *hot* outbursts of this symbiotic system. Moreover, the absorption component observed in the profiles of the emission lines He I (λ 6678), H$\alpha$ and H$\beta$ completely disappeared during this outburst, which again testifies to its *hot* character.

The EWs of emission lines H$\alpha$, H$\beta$, He I (λ 6678) and He II (λ 4686) manifest an even more prominent increase during the minor outburst G1. Such behaviour would suggest that this brightening belongs also to the *hot* outbursts of AG Dra. On the other hand, the EWs of the Raman-scattered O VI (λ 6825) line dropped to deep minimum during this outburst, which was observed only during the major, *cool* outbursts. By that manner, the outburst G1 manifested the spectroscopic behaviour of both *hot* and *cool* outbursts of AG Dra (Merc et al., 2017). The open question remains whether it is a new type of outburst or some kind of transition between (or combination of) the *hot* and *cool* outbursts?

Although the third outburst G2 during the recent active stage of AG Dra was similar to the pre-outburst G0 in its brightness, we detected the same prominent increase of all the emission line EWs as in the case of brightening G1. Only exception was the H$\alpha$ line: its EWs were comparable to ones during the G0 pre-outburst. Other interesting feature of this outburst was the weakening of hydrogen emission lines just before the G2 outburst. Overall spectroscopic behaviour ranks this brightening as the *hot* outburst of AG Dra.

The last outburst of the symbiotic system detected in April 2018 was also of the *hot* type. The maxima of the EWs were either comparable to the values reached during previous outbursts (in the case of H$\alpha$, He I (λ 6678) and Raman-scattered O VI (λ 6825) lines) or demonstrated the highest values detected during the ongoing active stage of AG Dra (H$\beta$ and He II (λ 4686) lines).

## Conclusions

Periodical outbursts and their relation to periodicities in the symbiotic system AG Dra have been a matter of long-term debate. As mentioned in the introduction, González-Riestra et al. (1999) have distinguished between *cool* and *hot* outbursts of AG Dra according to the spectroscopic behaviour of this interacting binary observed in the far ultraviolet. In our previous study (Leedjärv et al., 2016), we showed that *cool* and *hot* outbursts of AG Dra can be clearly distinguished by the behaviour of the emission lines in the optical spectrum of this symbiotic system.

To sum it up, photometric as well as spectroscopic behaviour suggests that the last four outbursts of AG Dra belong to the minor, *hot* type. Such classification is also supported by the results of our analysis of the hot component's temperature during the ongoing active stage of this symbiotic system (more details are given in Merc et al., 2019). On the other hand, some specific effects observed during the outburst G1 (e.g. the almost disappearance of the Raman-scattered O VI lines) are more typical for the *cool* outbursts, despite the fact that the WD's temperature reached the historical maximum during this event.

The future evolution of AG Dra is an open question. Can we expect (finally) a major, *cool* or (again) minor, *hot* outburst? Another possibility is, that the symbiotic system will return to quiescence as we have already detected such behaviour during the weak activity stage 1963–1966. According to our detailed period analysis of photometric and spectroscopic observations we know that the median of the time interval between outbursts is around 360 days. It is worth noting that these time intervals vary from 300–400 d without an apparent long-term trend. Nevertheless, we are able to predict the time of next outburst of AG Dra during the spring of 2019. We can expect it in the interval from JD 2 458 581 (April 7, 2019) to JD 2 458 625 (May 21, 2019).

In any case, AG Dra clearly demonstrates the importance of long-term monitoring of symbiotic stars in order to disentangle the nature and mechanisms of their active stages and outbursts.

## Acknowledgement

We acknowledge with thanks the variable star observations from the *AAVSO International Database* and *ARAS Database* contributed by observers worldwide and used in this research. This work was supported by the Slovak Research and Development Agency project APVV 15-0458, the Estonian Ministry of Education and Research institutional research funding IUT 40-1, the Russian Foundation for Basic Research project No. 17-52-45048 and European structural and investment funds and the Czech Ministry of Education, Youth and Sports project